\documentclass{article}
\usepackage{arxiv,amsmath,graphicx,url,times,booktabs, tabularx, amsfonts}

\title{A Proposal for Foley Sound Synthesis Challenge}

\twoauthors
  {Keunwoo Choi, Sangshin Oh, Minsung Kang}
    {Gaudio Lab, Inc.\\
     Seoul, South Korea\\
     keunwoo@gaudiolab.com}
  {Brian McFee
  }
    {New York University \\
     New York, USA\\
     brian.mcfee@nyu.edu}

\begin{document}

\ninept
\maketitle

\begin{sloppy}

\begin{abstract}



``Foley'' refers to sound effects that are added to multimedia during post-production to enhance its perceived acoustic properties, e.g., by simulating the sounds of footsteps, ambient environmental sounds, or visible objects on the screen.
While foley is traditionally produced by \emph{foley artists}, there is increasing interest in automatic or machine-assisted techniques building upon recent advances in sound synthesis and generative models.
To foster more participation in this growing research area, we propose a challenge for automatic foley synthesis.
Through case studies on successful previous challenges in audio and machine learning, we set the goals of the proposed challenge: rigorous, unified, and efficient evaluation of different foley synthesis systems, with an overarching goal of drawing active participation from the research community.
We outline the details and design considerations of a foley sound synthesis challenge, including task definition, dataset requirements, and evaluation criteria. 


\end{abstract}


\section{Introduction}
\label{sec:intro}

Foley, or foley sound, refers to the reproduction of everyday sound effects that are added to multimedia to enhance the audio quality \cite{stinson1999real}. In professional media production such as films and TV shows, the use of foley is widespread; including sound events (footsteps, gunshots, cars, crowds) and environmental effects (rain, wind, snow). To meet the demands for various acoustic events, post-production studios usually own a large sound effect catalog. To obtain a perfectly matched sound effect, it is common to edit the existing sound effects or record a new one. The recording process may need to be creative or even artistic. For example, a common trick of `Corn starch in a leather pouch makes the sound of snow crunching' is introduced in Wikipedia. As much as it may sound interesting, it is a challenging and tedious process, yet necessary to be done for many sound events in multimedia content. 

The benefit of foley sound synthesis technology is evident; to make the workflow more efficient. Recently, researchers started to apply deep neural networks to generate foley sounds, motivated by the recent success in the generation of speech and music signals~\cite{oord2016wavenet, wang2017tacotron}. Nevertheless, It is in its early days for the problem - there is no standard in problem definition, dataset, and evaluation. 

One effective way to make collective research progress is to establish a challenge. Setting a standard problem definition and evaluation schemes is already a success, which is naturally done during specifying a challenge. Challenges also stimulate research communities by rewarding successful research outcomes explicitly.




In this paper, we propose a foley sound synthesis challenge. Our goal is to discuss ideas about the proposed challenge, which is relevant to DCASE, and position this proposal to ultimately create an official challenge in the upcoming DCASE workshops. We review recent machine learning challenges in audio, speech, and music research in Section~\ref{sec:casestudy} and existing works and datasets in Section~\ref{sec:background}. In Section~\ref{sec:challenge}, we provide a proposal for foley sound synthesis challenge that includes problem definition, datasets, and evaluation metrics. We conclude the paper in Section~\ref{sec:conclusion}.

\section{Case Study: Research Challenges}\label{sec:casestudy}
In this section, we review five existing research challenges: Blizzard Challenge, CHiME, DCASE, Music Demixing challenge, and AI Song Contest. The former three are relatively mature while the latter two started after 2020. All of them started along with the increasing popularity of the research problems and have contributed to the continued growth by defining the tasks, providing common datasets, and performing reliable evaluation. 

The Blizzard Challenge started in 2005 ``in order to better understand different speech synthesis techniques on a common dataset''~\cite{black2005blizzard}. In their first challenge, the participants submitted synthesized speech on five text genres including novels and news. To focus on speech synthesis techniques and not on text analysis techniques, relatively simple texts were selected by the organizers. The results were evaluated by three categories of listeners -- speech experts, volunteers (`random' users), and US undergraduates. The evaluation was performed in a fully subjective manner. For simpler three genres, the listeners were asked to rate them on a 5-point scale; while for the other two genres, they were to transcribe what they hear so that the speech intelligibility can be assessed. It remains to be an important annual event in speech synthesis research until now.

The CHiME challenge on speech separation and recognition was proposed and started in 2011 as a refined version of previous, similar challenges~\cite{kolossa2011chime, barker2013pascal}. The refinement includes modernization of the problem definition (types of noise, signal mixing models, etc), adjustment of the difficulty, and introduction of more realistic datasets/evaluation metrics. For researchers, CHiME remains to be one of the most important venues where novel methods are introduced and tested on their rigorously created datasets.



Detection and Classification of Acoustic Scenes and Events (DCASE) is one of the most active and successful challenges in the field of audio research~\cite{giannoulis2013detection, stowell2015detection}.
As a part of IEEE AASP workshop, the first DCASE hosted two tasks -- acoustic scene classification and sound event detection -- and accepted 21 systems. Six papers were presented in the DCASE poster session in WASPAA 2013 later. In 2016, DCASE started to host is own workshop along with the challenge. In 2021, DCASE had 394 submissions for six tasks and is recognized as the most relevant and prominent academic venue in acoustic scene analysis. The scope of DCASE has been expanded towards some highly practical applications (low-complexity acoustic scene classification and anomaly detection for machine monitoring), bioacoustics (mammals and bird sound understanding), and natural language understanding (audio captioning and natural language queries).

Demixing challenge and AI Song Contest are founded 2021 and 2020 \cite{mitsufuji2021music, huang2020ai}. In Demixing challenge, participants submit their music source separation systems with specifying whether external datasets were used or not. Although objective metrics in source separation have limits in measuring perceptual quality, an objective metric (signal-to-distortion ratio) was used to evaluate the systems since it is known to represent the performance of source separation fairly well. Conversely, in AI Song Contest, participants submit final tracks which are then evaluated purely subjectively -- by judges and online votes. Given the artistic aspect of the task, it is a reasonable choice, although the details such as the balance between judges and public votes may be subject to change.


\section{Backgrounds}\label{sec:background}

\subsection{Previous Works on Foley Sound Synthesis} \label{subsec:prevw}


Foley sound synthesis (FSS) has been studied using traditional audio synthesis approaches \cite{hahn1998integrating, cook2002modeling, fontana2003physics, turchet2016footstep}. 
The system introduced in \cite{hahn1998integrating} generates sounds that are synchronized to motions in a virtual environment. In \cite{cook2002modeling, fontana2003physics, turchet2016footstep}, researchers focused on generating more realistic footstep sounds.


In the more recent, data-driven approaches, the target sound categories have expanded significantly by researchers taking advantage of deep generative models such as GANs and WaveNet~\cite{goodfellow2014generative, oord2016wavenet, wang2017tacotron}.
The authors of \cite{owens2016visually} proposed the first deep learning based system that generates the sounds in a video dataset, `The Greatest Hits', which was also introduced in the same paper. 
This dataset consists of recordings of special and artificial actions - hitting and scratching a drum stick on various types of surfaces. 
Regarding the limited variety of actions and acoustic events, the proposed system is a solution to a simplified foley generation. Still, the system needs to learn the type and timing of the visual events and generate a relevant audio signal.   

There are also systems that generate a wider range of foley sounds, for example, fireworks, dog barking, footsteps, gunshots, etc. These works are based on more realistic datasets, such as AudioSet, VEGAS, and VGGSound \cite{gemmeke2017audio, zhou2018visual, chen2020vggsound}, that consist of videos in the wild. For generating video-synchronized FSS, GANs \cite{liu2021towards, chen2020generating, ghose2021foleygan, iashin2021taming} have been a popular method. 
%
They all have a similar concept of generative model being conditioned by a visual encoder to synthesize STFTs or waveforms. But there are some differences in the input types of the sound generation models. 
In \cite{liu2021towards, chen2020generating}, the sound generator modules are conditioned by an embedding of the input video.
Particularly in \cite{chen2020generating}, the sound generator is conditioned by a visual feature that is co-trained with the corresponding audio feature, so that the visual model can learn to focus on visual objects that are relevant to audio. 
In \cite{ghose2021foleygan}, the authors used the visual action class and the visual action spectrogram to condition the sound generator.
In \cite{iashin2021taming}, the authors used a VQ-GAN model which is a combination of VQ-VAE and an adversarial loss. 

%


%

Majorities of the aforementioned works are based on visual queries, making the system follow the workflow of foley artists. 
As a consequence, FSS systems are often required to have a visual understanding module.
In \cite{guzhov2022audioclip, wu2022wav2clip}, for example, three modalities -- text, vision, and audio -- are combined using pre-trained CLIP encoders \cite{radford2021learning}, and these encoders are utilized for subtasks like classification, retrieval, etc.


Based on the recent progress in models and datasets, there is strong evidence for us to believe that the foley sound synthesis is at its tipping point. There are many powerful audio generative models, e.g., auto-regressive models~\cite{iashin2021taming}, VAEs~\cite{dhariwal2020jukebox, iashin2021taming}, GANs~\cite{ghose2021foleygan, liu2021towards}, and diffusion-based models~\cite{kong2020diffwave}. Visual understanding models are mature enough to be used in various applications. Moreover, there are audio-visual datasets that have rich information as we will review more deeply in the next section.

\subsection{Existing Datasets}\label{subsec:prevd}
In this section, we first review four video datasets -- AudioSet~\cite{gemmeke2017audio}, VEGAS~\cite{zhou2018visual}, VGGSound \cite{chen2020vggsound} and The Greatest Hits~\cite{owens2016visually}. These datasets have been used in the previous FSS works~\cite{owens2016visually, zhou2018visual, ghose2020autofoley, ghose2021foleygan, chen2020generating}. We also review three more datasets that are based on Freesound.org, an online sound database: UrbanSound8k~\cite{Salamon:UrbanSound:ACMMM:14}, FSDKaggle2018~\cite{eduardo_fonseca_2019_2552860}, and Clotho~\cite{konstantinos_drossos_2021_4783391}. 

Video datasets have been the main source of data in the previous FSS research. \textbf{AudioSet}~\cite{gemmeke2017audio} is a 5800-hour video dataset from YouTube. It consists of 10-second clips which are manually labeled with 527 audio event classes. Since all the videos are sampled from YouTube, low audio and video quality can be a problem.
%
\textbf{VEGAS}~\cite{zhou2018visual} is a curated subset of AudioSet, 55-hour in total and 7-second on average. Its creators selected 10 classes and filtered the items to include videos that have a direct audio-visual relationship only.
\textbf{VGGSound}~\cite{chen2020vggsound} has more than 550 hours of 10-second video clips that cover 310 classes of sound events. Similar to VEGAS, the creators carefully selected samples to remove videos that have a weak audio-visual relationship. 
\textbf{The Greatest Hits} \cite{owens2016visually} is a dataset that consists of action-sound pair videos, hitting and scratching a drum stick on various surfaces, as explained in Section~\ref{subsec:prevw}.

Audio datasets without visual information can be also a good source of data for the FSS task. 
\textbf{Urbansound8k}~\cite{Salamon:UrbanSound:ACMMM:14} is a sound dataset that includes 8,732 short audio segments. It has been a useful source for the acoustic event classification task. However, the items in this dataset are 4-second segments that were trimmed from longer audio files, which reduces the overall sound diversity of the dataset.
\textbf{FSDKaggle2018}~\cite{eduardo_fonseca_2019_2552860} contains 11,073 audio files that are labeled with 41 classes, following the AudioSet ontology. The items were sampled from Freesound and then manually annotated on MTurk. 
\textbf{Clotho}~\cite{konstantinos_drossos_2021_4783391} is a dataset that was created for the audio captioning task. Each audio file has 5~captions; i.e., its 4,981 audio files are provided with 24,905 audio captions. Audio samples are 15 to 30 second in duration and captions are 8 to 20 words. This can be potentially used to develop a text-query FSS system.
%
%




None of the existing datasets provides clean audio samples as desired when using foley sounds. It may be possible to generate high-quality foley sounds using noisy sounds. However, in the initial version of FSS challenge, we believe it would be better to have a dataset with high quality foley sounds to simplify the task. 



\section{Foley Sound Synthesis Challenge}\label{sec:challenge}

\subsection{Consideration on FSS Problem Definition}
A FSS problem can be defined by specifying several aspects. In this section, we first review those aspects, which are followed by the details of our proposal. We propose to initiate the challenge with a narrow and strict version while leaving definitions of more complicated systems for the future.

\textbf{Type of sound} - Foley can be categorized into two classes: i)~one that belongs to a single acoustic event (e.g., gunshot) or ii)~one that determines the atmosphere/ambience of a scene (e.g., rain or room tone). 

\textbf{Reality of acoustic event} - Some foley sounds are bound to realistic objects or acoustic events, e.g., honks of existing car models, dog barks, door closing sounds, etc. Others are rather outcomes of human imagination, e.g., laser guns or dinosaurs. Due to the lack of reference, the generation of the sound of the latter is rather an art, making it difficult to evaluate either objectively or subjectively. 

\textbf{Type of input query} - It may seem natural for FSS systems to take video as input as human sound engineers and foley artists do. However, there is a drawback if the FSS problem is formulated in such a way. Automatic video-to-sound generation is a combination of multiple machine learning problems that include not only audio synthesis but also video understanding. This would make evaluation more complicated than audio quality assessment, which is our primary focus. A simpler alternative is text-to-sound where the text may be a description of a scene or a sound. Even further, the simplest problem formulation is perhaps a system that takes a sound category as input (e.g. a one-hot-vector). 


\textbf{Final audio format} - The simplest form of generated foley sound would be an anechoic single-channel audio signal. One can also consider adding more audio processing such as multi-channel mixing and reverberation.

\subsection{Our Suggestion: Progressive Approach}\label{subsec:define_problem}
After taking those considerations into account, we suggest a progressive approach as described below, starting from the simplest problem definition and moving towards complicated ones as the challenge evolves over years. 





\begin{description}

\item[Level 1: Categorical sound generation]
The most simplified form of the foley generation problem consists of generating individual sound clips from categorical identifiers (\emph{footsteps}, \emph{rain}, etc).
This formulation is useful for constructing or expanding a sample library from which foley artists could select clips to use in a particular scene.
Systems would be evaluated by (perceptual) quality of generated audio and diversity of generated samples for each category.

\item[Level 2: Sequential generation from text description]
Expanding from level 1, this formulation would provide a textual description of the scene as input, from which systems would generate corresponding audio clips.
This is intended to model both the clip generation task from level 1 and the composition of individual clips into a plausible acoustic scene.
Each textual prompt would contain one or more sound events, and systems would be evaluated by the criteria of level 1, and by relevance and coverage in response to the prompt.

\item[Level 3: Sequential generation from video]
The next more advanced formulation would work directly from the video sequence, bypassing categorical and textual description entirely.
This can be seen as a minimal `fully automated' foley system in the sense that no manual intervention is necessary to describe or compose the scene.
Systems would be evaluated along similar criteria to level 2.

\item[Level 4: Mixed soundtrack generation]
The final level expands upon level 3 by incorporating multi-channel mixing (stereo,
5.1, etc.) into the generation process.
In addition to the evaluation criteria of previous levels, subjective measures of
stereo separation and spatial immersion will be included.

\end{description}

\subsection{Consideration on the Official Dataset}



\textbf{Simulated sound} - The foley is often an audio signal that is recorded and edited to give a certain acoustic impression, regardless of how an actual acoustic event would sound. For example, a good foley of `body hitting' would be a loud and exaggerated version of low-frequency impulsive sounds rather than an actual recording of body hitting which is barely audible. 


\textbf{Accompanied information} - As we discussed in Section~\ref{subsec:prevd}, the dataset may or may not have visual cues. This will depend on how the problem is formulated.  

\subsection{Our Suggestion: Official Dataset}
In this section, we propose the dataset specification that is compatible with the FSS problem definition at level 1 in Section~\ref{subsec:define_problem}.
In Gaudio Lab, where some of the authors are affiliated, there are in-house foley artists and the company is willing to provide audio files to create the official dataset.

\textbf{Number of category} - In the same spirit of narrowing down the problem in level 1 definition, the number of categories would be limited to a small number. Still, the categories should be chosen to cover diverse foley sounds. For example, \textit{four} categories can be selected; footsteps and gunshots (impulsive sounds), dog barking (tonal sounds), and swooshes (airborne sounds with a varying duration).

\textbf{Number of item} - It can be critical to the performance of the FSS systems. We assume that the number of items should be larger than a hundred per category but this is subject to change and would need more discussion in the future.

\textbf{Audio quality} - full-bandwidth (44.1~kHz) signals would be desired considering the users of foley sound, which are, professional content creators. 

\textbf{Visual cue} - Following the problem definition of level 1 in Section~\ref{subsec:define_problem}, in the first version of the official dataset, there would not be visual information. 


  
%

\subsection{Consideration on Evaluation}

Evaluation of generative models or synthesized samples is a difficult task in general, yet very important particularly when organizing a challenge. Objective evaluation is convenient, but it often disagrees with human perception. It may be also `gamed', which would be undesirable for a challenge.
As an alternative, human-engaged evaluation has been widely used as a way of benchmarking to compare performance for different audio generation models~\cite{hayashi2020espnet}.
In practice, however, subjective evaluation may not be preferred in a large-scale evaluation due to the low time- and cost-efficiency. 
In this section, we discuss the objective and subjective metrics to evaluate generative models.

\subsubsection{Objective Evaluation Metrics}

    \textbf{Inception score (IS)} 
    has gained popularity after its introduction to evaluate implicit generative models such as GANs. IS is defined with the posterior distribution of auxiliary classifier models.
    %
    IS is often interpreted as multiplication of sharpness and diversity, which represent the confidence of the classifier in its output labels and the class diversity of the generated samples, respectively.
    In other words, IS is able to sensitively reflect the quality of class identity and diversity. However, it does not capture the mode inconsistency, i.e. mode addition or dropping.

    \textbf{Fréchet inception distance (FID)} 
    is another widely used metric for generative models. Instead of using \textit{outputs} of an auxiliary model as done in IS, the calculation of FID relies on \textit{hidden representations} of an auxiliary model. The auxiliary model does not need to be a classifier, and it is assumed that FID is robust to choice of representation space. Recently, FID was adopted for audio generation using an audio classifier and named Fréchet audio distance (FAD)~\cite{kilgour2019frechet}. 
    
    %
    FID (as well as FAD) is calculated from sets of hidden representations of generated and real samples. Each set of representation is fitted to a multivariate Gaussian distribution, and the Fréchet distance of these two distributions becomes the FID score of the generated samples.
    Since it directly models the distribution of hidden representations, FID is sensitive to mode changes including spurious mode addition or dropping \cite{sajjadi2018assessing}.

\textbf{Likelihood} is worth discussing, although we do not recommend using it. It is
    one of the most straightforward metric for evaluation of generative models. For density estimation, likelihood is a key metric as it directly shows how the model estimates the target distribution. However, we suggest not using likelihood for two reasons. First, it is often not aligned with perceptual quality \cite{Theis2016a}. Second, it has limited applicability since calculation of likelihood is not possible for some generative models such as including GANs, VAEs or diffusion probabilistic models. 
    
\subsubsection{Subjective Evaluation Metrics}
Perhaps the most common scheme of subjective evaluation is to ask the mean opinion score (MOS). In a MOS test, human raters are guided to evaluate a sample into five-point scale where 1 means \textit{very bad} and 5 means \textit{very natural}. Usually, there are various aspects that affect the overall perceptual quality such as fidelity, degree of artifacts, and naturalness. To measure them separately, it is common to design multiple questions for each sample.

\subsection{Our Suggestion: Two-Phase Evaluation}

To combine advantages of objective and subjective evaluation, we suggest a two-phase evaluation scheme as follows:
\begin{itemize}
    \item
        \textbf{Phase 1}: All participants will be evaluated only with objective metrics. All metrics will be publicly announced and considered simultaneously in the public leader board.
        %
        %
        The participating systems will be scored with respect to each objective metric. The scores will be standardized over all the evaluators, and the standardized scores will be summed to yield the final score of a system.
    \item
        \textbf{Phase 2}: Only the top-$N$ submissions from phase 1 will be evaluated on five-point scale (MOS). Human evaluators will be asked to focus on various aspects of the generated samples, e.g., fidelity, coincidence with class label, and temporal alignment. The final ranking will be decided by an average MOS.
\end{itemize}

In the first phase, inception score (IS) and Fréchet inception distance (FID) will be used to determine ranks of the participants. In order to detect severe memorization of training samples, which can happen as a result of overfitting, we will use the memorization-aware version of FID \cite{bai2021training}.

In the second phase, we suggest the following criteria in the subjective evaluation.
\begin{itemize}
    \item Fidelity of generated audio
    \item Distortions and artifacts of generated audio
    \item Coincidence to the given conditions, e.g. class labels, text descriptions, or video frames
\end{itemize}

The evaluators do not need to be an experienced audio engineers. However, it will be important to ensure that the quality of the testing environment (headphones and the listening room) meets some criteria.

\section{Conclusion}\label{sec:conclusion}
In this paper, we proposed a new challenge in audio research -- foley sound synthesis. We reviewed several recent challenges including their goals, progresses, and impacts. We also reviewed the existing research and datasets in foley sound synthesis. Based on these reviews, we presented a brief proposal on foley sound synthesis challenge including our suggestion on problem definition, datasets, and evaluation. We hope this paper triggers discussion that leads to initiation of the proposed challenge, and ultimately, progress in foley sound synthesis.




\bibliographystyle{IEEEtran}
\bibliography{refs}

\end{sloppy}
\end{document}